\documentclass[prl, twocolumn, superscriptaddress]{revtex4}
\usepackage{graphicx,color}
\usepackage{amssymb}
\usepackage{amsmath}
\usepackage{bm}
\usepackage[colorlinks=true,linkcolor=blue,urlcolor=blue,citecolor=green]{hyperref}
\usepackage{tabularx}
\usepackage{multirow}
\usepackage{braket}
\usepackage{xcolor}
\usepackage{gensymb}
\usepackage{amsmath,amssymb}

\begin{document}

\title{Ferroelectric Hysteresis in Superconducting Bilayer T$_{\textrm{d}}$-MoTe$_2$}
\author{Yanfang Li}\thanks{These authors contributed equally to this work}
\affiliation{State Key Laboratory of Quantum Functional Materials, School of Physical Science and Technology, ShanghaiTech University, Shanghai 201210, China}
\author{Xin-Zhi Li}\thanks{These authors contributed equally to this work}
\affiliation{State Key Laboratory of Quantum Functional Materials, School of Physical Science and Technology, ShanghaiTech University, Shanghai 201210, China}
\author{Wen-Yu He}\thanks{hewy@shanghaitech.edu.cn}
\affiliation{State Key Laboratory of Quantum Functional Materials, School of Physical Science and Technology, ShanghaiTech University, Shanghai 201210, China}
\affiliation{\mbox{ShanghaiTech Laboratory for Topological Physics, ShanghaiTech University, Shanghai 201210, China}}

\date{\today}

\begin{abstract}
Recently, the displacement-field-driven hysteretic switching of superconductivity was reported in ferroelectric bilayer T$_{\textrm{d}}$-MoTe$_2$. Such direct coupling between ferroelectricity and superconductivity offers promising pathways for low-power, non-volatile memory devices, but the underlying coupling mechanism remains poorly understood. Here, we demonstrate that the ferroelectric switching of superconductivity can naturally originate from an intralayer, $p$-$d$ orbital pairing. In bilayer T$_{\textrm{d}}$-MoTe$_2$, the ferroelectric polarization segregates the $p$- and $d$-orbital electrons into distinct layers, thereby suppressing the intralayer, $p$-$d$ orbital pairing. By developing a phenomenological Landau-Ginzburg model, we establish that the hysteretic switching of superconductivity requires $P_{\textrm{r}} < P_{\textrm{c}} < P_{\textrm{s}}$, where $P_{\textrm{c}}$ is the critical pair-breaking polarization and $P_{\textrm{r}}$ ($P_{\textrm{s}}$) is the remanent (saturated) polarization. Crucially, our scenario of intralayer, $p$-$d$ orbital pairing indicates that the bilayer T$_{\textrm{d}}$-MoTe$_2$ features an anisotropic momentum-dependent pairing gap and can transition into a pair density wave by tuning the chemical potential, which provides clear pathways for experimental verification.
\end{abstract}

\maketitle

\emph{Introduction}.--- Atomically thin two-dimensional (2D) ferroelectrics are appealing for next-generation non-volatile memories~\cite{Uchino, Rappe, Khan} and have generated significant scientific interest~\cite{Menghao1, Menghao2, BTZhou, Bauer, Hongjun, Tsymbal, Chuanshou, Seidel00}. Recently, a number of 2D van der Waals layered materials have been found to exhibit ferroelectricity~\cite{Chuanshou, Seidel00, Zaiyao, Seidel, Jianhua, Pablo1, Pablo2, Shalom, Hunt, Gorbachev, Pablo3, Fucai, Yiwan, Jianming, Junhao, Bian01, Jianming2, Guangyu01, Yasuda001, Shuigang, LanChen, Rhodes, Zizhong, Pablo4}. In these 2D ferroelectric materials, stacking configurations break the inversion symmetry and lead to electrically switchable polarizations~\cite{Chuanshou, Seidel00}. Notably, the hallmark of the ferroelectric order in these 2D materials is an electric polarization hysteresis loop driven by an out-of-plane electric field~\cite{Chuanshou, Seidel00, Seidel, Jianhua,Yiwan, Shalom,littlewood}.

For ferroelectrics in 2D metals, the ferroelectric hysteresis of electric polarization is manifested as a resistance hysteresis during sweeps of gate voltage~\cite{Zaiyao, Pablo1, Pablo2, Hunt, Pablo3,Jianming}. This gate-induced resistance hysteresis has been observed in bilayer T$_{\textrm{d}}$-MoTe$_2$~\cite{Rhodes, Zizhong}, establishing the bilayer T$_{\textrm{d}}$-MoTe$_2$ as a ferroelectric metal. Most strikingly, the ferroelectric hysteresis of resistance in bilayer T$_{\textrm{d}}$-MoTe$_2$ persists from the normal state into the superconducting state~\cite{Rhodes, Zizhong}. Moreover, similar ferroelectric hysteresis has been independently observed in superconducting twisted bilayer graphene~\cite{Pablo4}. These parallel observations in two distinct superconducting bilayer systems provide compelling evidence for a direct coupling between ferroelectricity and superconductivity.

The unprecedented coupled ferroelectricity and superconductivity in bilayer T$_{\textrm{d}}$-MoTe$_2$ and twisted bilayer graphene profoundly challenges the long-standing paradigm that ferroelectricity and superconductivity are mutually exclusive~\cite{Yasuda01, Ariando, Zhenyu01}. As the conventional incompatibility between ferroelectricity and superconductivity has been circumvented in these superconducting bilayers, a fundamental question arises: what is the microscopic mechanism behind the intriguing coupling between the ferroelectricity and superconductivity?

Experimentally, the hysteresis between the resistive and superconducting states reveals a ferroelectric switching of superconductivity: an enhanced polarization induced by a large gate voltage destroys superconductivity, whereas the superconducting state coexists with the remanent polarization $\left(P_{\textrm{r}}\right)$ at zero displacement field. In bilayer T$_{\textrm{d}}$-MoTe$_2$, this antagonistic interplay is most strikingly evidenced by the dramatic enhancement of the superconducting critical temperature $T_{\textrm{c}}$ just before the coercive reversal of the polarization~\cite{Rhodes}. Since the polarization usually decreases before reversal, these observations consistently indicate that the out-of-plane polarization is detrimental to the superconducting pairing in bilayer T$_{\textrm{d}}$-MoTe$_2$. Furthermore, given that monolayer T$_{\textrm{d}}$-MoTe$_2$ hosts a robust intrinsic superconducting state~\cite{Daniel-Nano}, it naturally follows that the superconducting pairing in the bilayer is predominantly intralayer in nature. Consequently, to unveil the mechanism underlying the ferroelectric switching of superconductivity, the central theoretical challenge is to identify what specific type of intralayer pairing is inherently suppressed by an out-of-plane ferroelectric polarization.

\begin{figure}[t]
  \centering
  \includegraphics[width=0.49\textwidth]{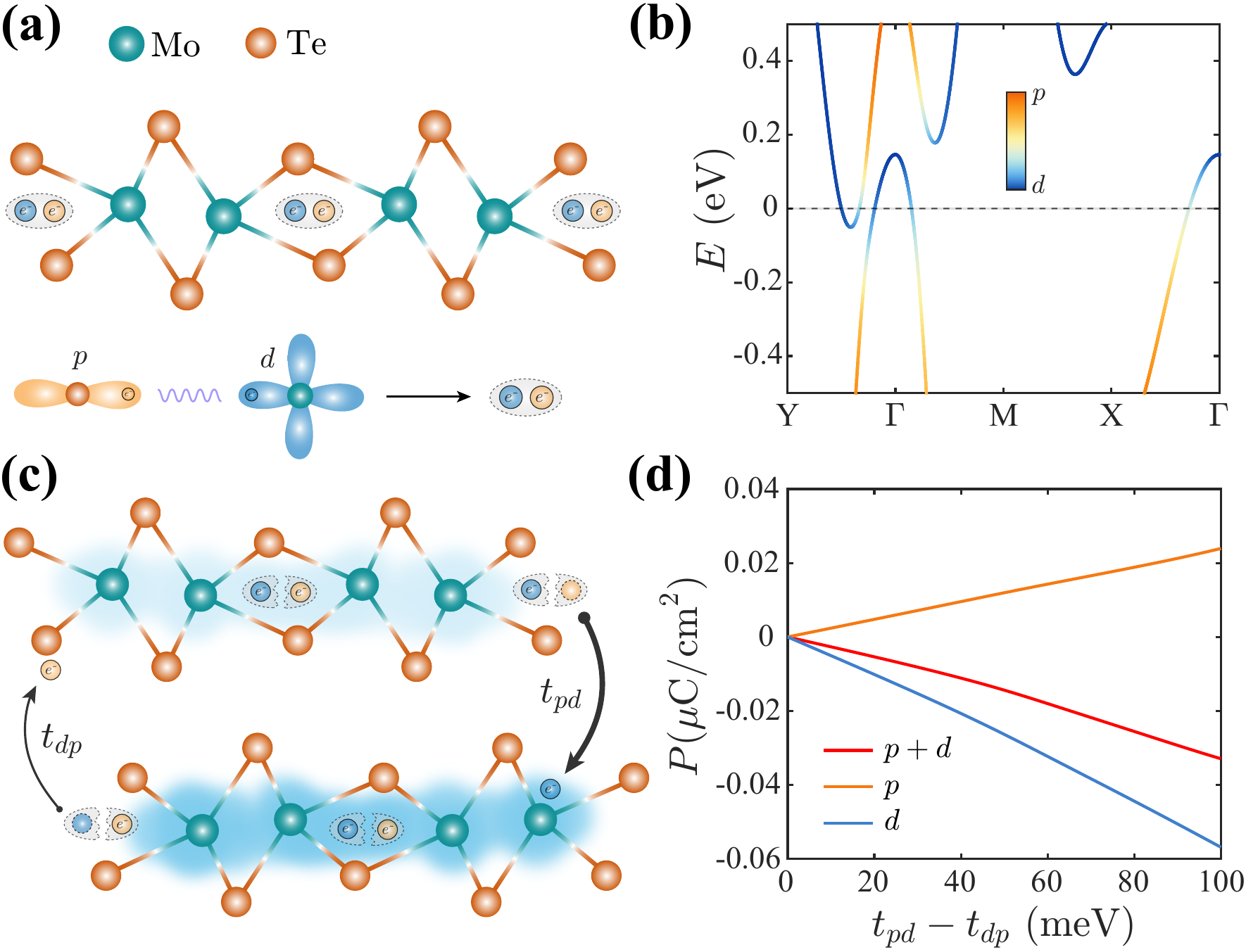}
  \caption{Stacking-induced ferroelectric polarization in bilayer 	T$_{\textrm{d}}$-MoTe$_2$. (a) Crystal structure of the 1T'-monolayer MoTe$_2$. The low-energy electronic states are dominated by Te $p_y$ and Mo $d_{yz}$ orbitals, and they form inter-orbital Cooper pairs. (b) Orbital-resolved low-energy band structure of the 1T'-monolayer. Spin-orbit coupling strongly hybridizes the electron-like $d_{yz}$ band and hole-like $p_y$ band near the Fermi energy. (c) The T$_{\textrm{d}}$ stacking of bilayer MoTe$_2$. The interlayer coupling asymmetry ($t_{pd}\neq t_{dp}$) polarizes the $p$- and $d$-orbital electrons into opposite layers, spontaneously generating an out-of-plane ferroelectric polarization that suppresses the intralayer $p$-$d$ orbital pairing. (d) Net layer polarization of electronic states driven by asymmetric interlayer hoppings $t_{pd}\neq t_{dp}$.}
  \label{fig1}
\end{figure}

\begin{figure}[t]
  \centering
  \includegraphics[width=0.49\textwidth]{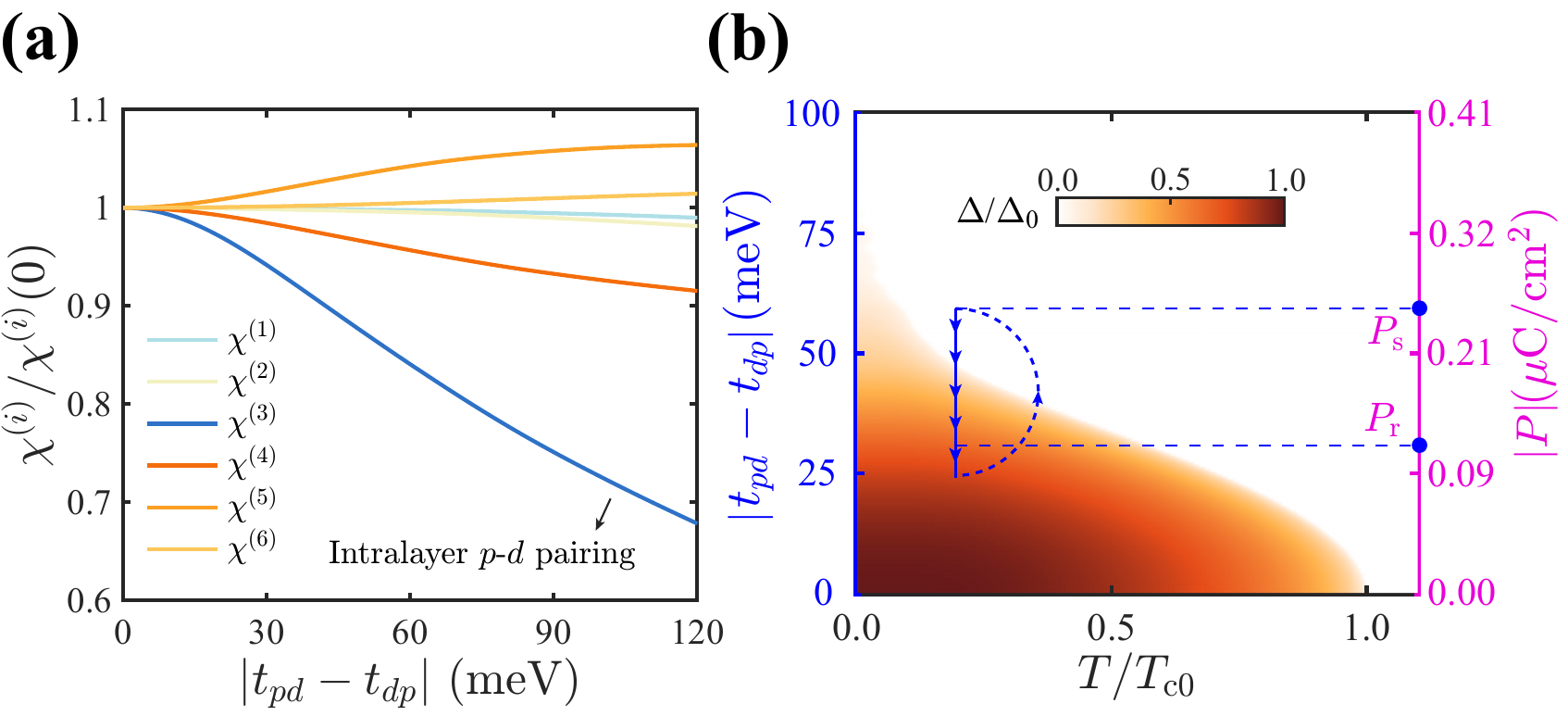}
  \caption{Ferroelectric suppression of intralayer $p$-$d$ orbital pairing. (a) Pairing susceptibilities versus interlayer coupling asymmetry $|t_{pd}-t_{dp}|$ for the six pairing channels ($\chi^{\left(i\right)}$ with $i=1,2,3,4,5,6$) allowed by the intralayer electron density-density interaction. The susceptibility in the intralayer $p$-$d$ orbital channel ($\chi^{\left(3\right)}$) is seen to get strongly suppressed by $|t_{pd}-t_{dp}|$. (b) Superconducting phase diagram obtained by minimizing the free energy. The intralayer $p$-$d$ pairing order parameter $\Delta$ decreases with increasing $|t_{pd}-t_{dp}|$ and vanishes at the critical polarization $P_{\textrm{c}}$. The blue trajectory in the diagram manifests the ferroelectric switching of superconductivity induced by the hysteretic exceeding of $P_{\textrm{c}}$.}
  \label{fig2}
\end{figure}

\begin{figure}[t]
  \centering
  \includegraphics[width=0.49\textwidth]{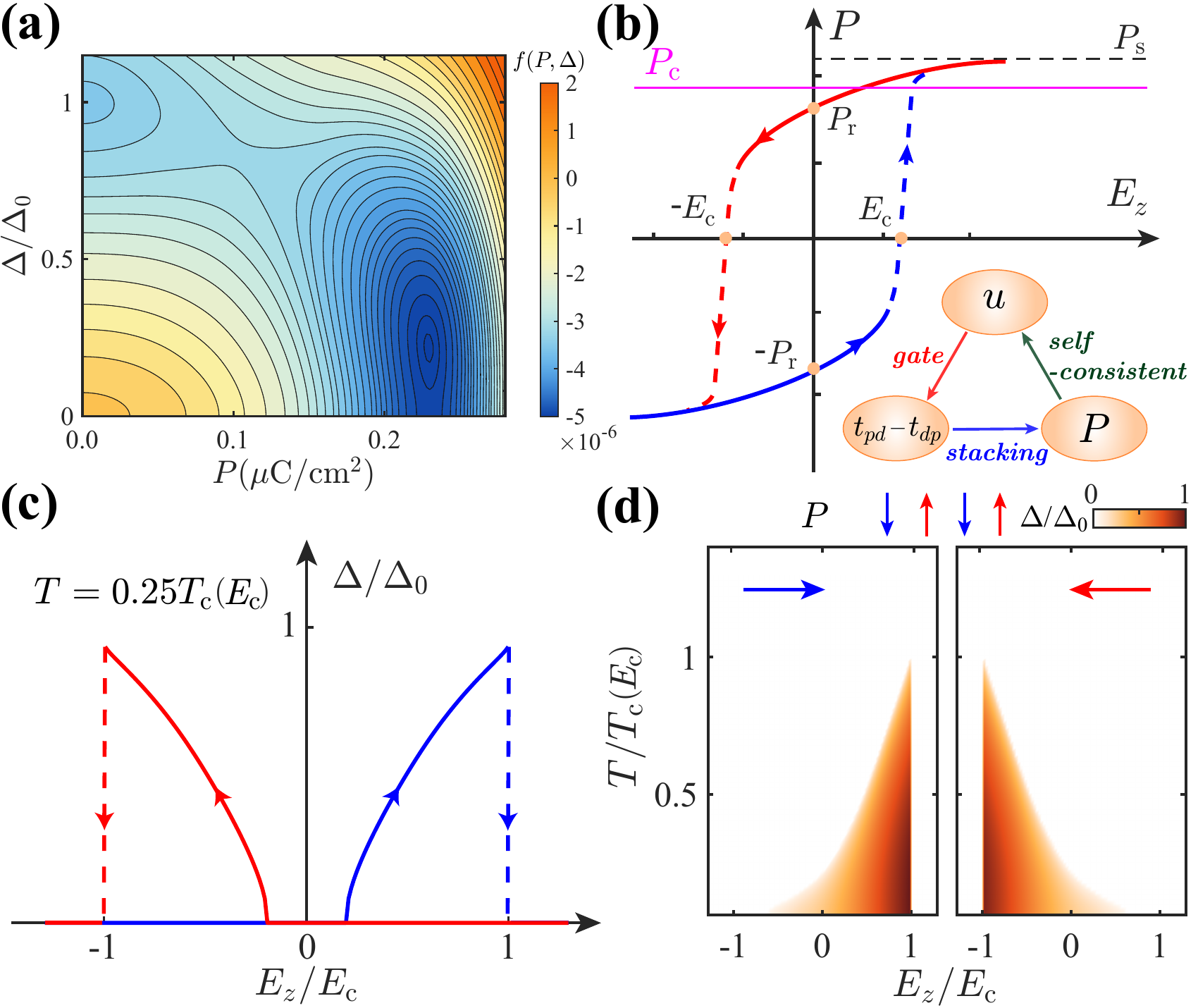}
  \caption{Hysteretic coupling between ferroelectricity and superconductivity. (a) Free energy landscape $f(P,\Delta)$ at zero external electric field. The global minimum at finite $P$ and $\Delta$ demonstrates the coexistence of ferroelectric polarization and intralayer $p$-$d$ pairing. (b) The ferroelectric polarization hysteresis under a cyclic out-of-plane electric field $E_z$. The superconducting state, coexisting with the remanent polarization $P_{\textrm{r}}$ at $E_z=0$, is destroyed when $P$ abruptly exceeds $P_{\textrm{c}}$ at $|E_z|>E_{\textrm{c}}$. The bottom-right inset schematically shows the self-consistent feedback loop among $u$ (controlled by the external gate voltage), the interlayer coupling asymmetry ($t_{pd}-t_{dp}$) and the ferroelectric polarization ($P$). (c) The corresponding hysteresis of the pairing order parameter $\Delta$. The switching of $\Delta$ between zero and a finite value directly manifests the resistance hysteresis that switches between finite and zero values. (d) Simulated temperature evolution of the coupled ferroelectricity and superconductivity. A dramatic increase in $T_{\textrm{c}}$ is observed as the ferroelectric polarization approaches reversal (indicated by arrows on top).}
  \label{fig3}
\end{figure}

In this work, we demonstrate that the ferroelectric hysteretic switching of superconductivity in bilayer T$_{\textrm{d}}$-MoTe$_2$ naturally emerges from an intralayer, inter-orbital pairing between the $p$ and $d$ orbitals. In the normal state, the T$_{\textrm{d}}$ stacking geometry imposes asymmetric interlayer couplings ($t_{pd}\neq t_{dp}$) that drive the $p$- and $d$-orbital electrons to polarize into opposite layers, thereby generating a spontaneous out-of-plane ferroelectric polarization (Fig.~\ref{fig1}). In the superconducting state, the intralayer, $p$-$d$ orbital pairing relies critically on the spatial overlap of the hybridized $p$ and $d$ orbitals. Consequently, by separating the $p$- and $d$-orbital electrons into distinct layers, the out-of-plane polarization induces strong pair breaking, yielding a critical polarization $P_{\textrm{c}}$ that completely destroys superconductivity (Fig. \ref{fig1} and \ref{fig2}). As a direct consequence of this polarization-induced pair-breaking mechanism, the ferroelectric hysteretic switching of superconductivity is shown to be governed by the condition $P_{\textrm{r}}<P_{\textrm{c}}<P_{\textrm{s}}$, where $P_{\textrm{s}}$ denotes the saturated ferroelectric polarization (Fig.~\ref{fig3}).

\begin{figure}[t]
  \centering
  \includegraphics[width=0.49\textwidth]{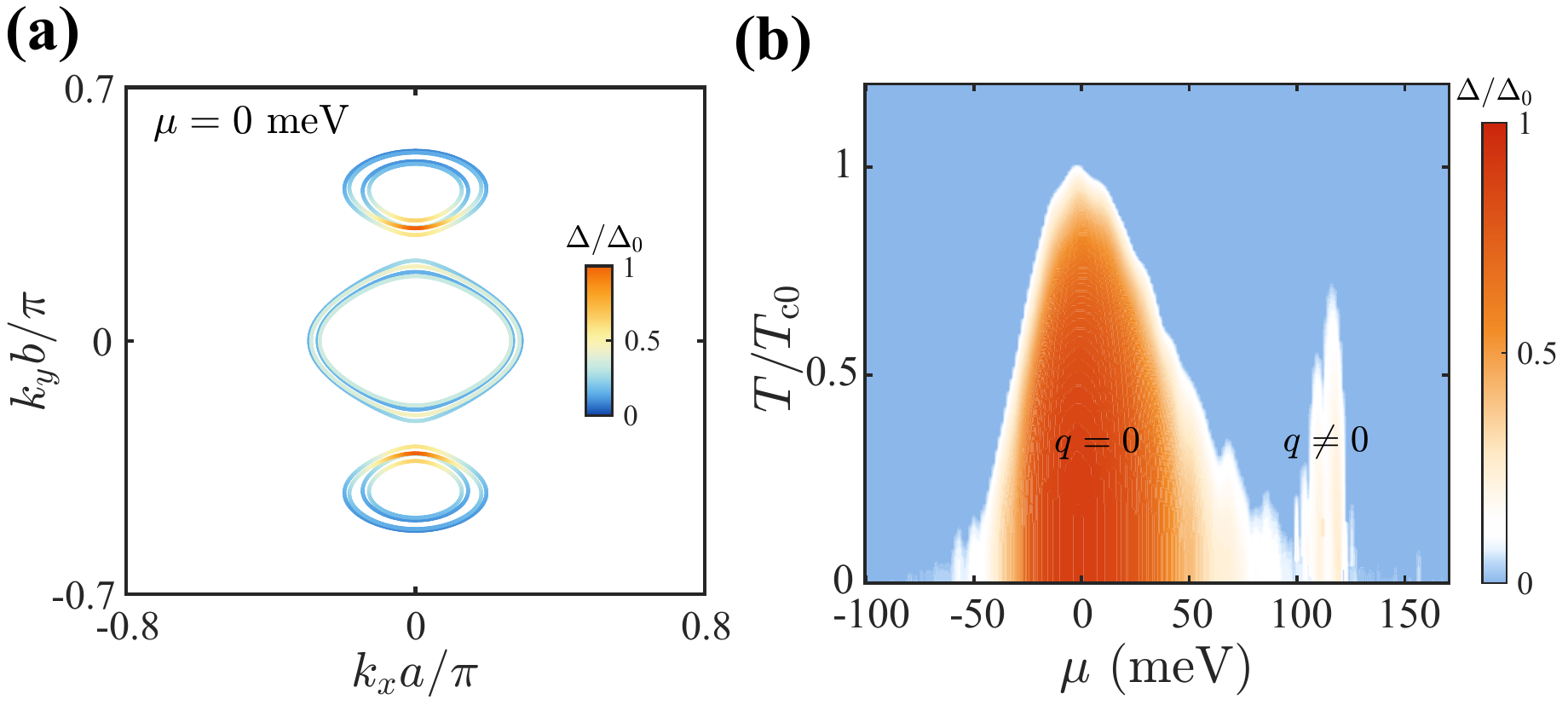}
  \caption{Anisotropic momentum-dependent pairing gap of the intralayer $p$-$d$ orbital pairing and the gate-induced PDW state. (a) The intralayer $p$-$d$ orbital pairing gap on the Fermi surfaces. The anisotropic pairing gap peaks in the region where the $p$- and $d$-orbitals strongly hybridize. (b) The $\mu$-$T$ phase diagram of the superconducting state. The transition temperature reaches its maximum at $\mu=0$ meV, where the $p$- and $d$-orbitals have the optimal hybridization, and decreases as the chemical potential shifts. Around $\mu=110$ meV, the Fermi pockets of $p$- and $d$-orbitals become well separated in the $\bm{k}$ space, so the intralayer $p$-$d$ orbital pairing interaction stabilizes a PDW state.}
  \label{fig4}
\end{figure}

In the following, we first analyze the microscopic origin of the stacking-induced ferroelectric polarization and explicitly elucidate how the out-of-plane polarization suppresses the intralayer, $p$-$d$ orbital pairing. Then, building upon this pair-breaking mechanism, we develop a Landau-Ginzburg model of the coupled ferroelectric order and intralayer, $p$-$d$ orbital pairing, showing that the hysteretic switching of superconductivity occurs as the polarization hysteretically exceeds $P_{\textrm{c}}$ (Fig.~\ref{fig3}). Finally, we discuss the physical implications of this intralayer, $p$-$d$ orbital pairing, proposing that the detection of an anisotropic momentum-dependent pairing gap and a gate-tunable pair density wave (PDW) state (Fig.~\ref{fig4}) can experimentally validate our scenario.

\emph{Stacking-induced ferroelectric polarization in the bilayer T$_{\textrm{d}}$-MoTe$_2$}.--- Our analysis starts with constructing the effective Hamiltonian for bilayer T$_{\textrm{d}}$-MoTe$_2$. The bilayer T$_{\textrm{d}}$-structure is formed by stacking two 1T'-monolayers related by a C$_{2z}$ rotation and a lateral half-unit-cell shift along the $y$ direction~\cite{BTZhou, Muechler}. In the 1T'-monolayer MoTe$_2$, the electronic states near the Fermi energy are predominantly derived from the Mo $d_{yz}$ and Te $p_y$ orbitals (Fig. \ref{fig1}(a)), which form the electron-like and hole-like bands: $E_{d,\bm{k}}$ and $E_{p,\bm{k}}$, respectively~\cite{Muechler,Xinzhi, Guangtong}. While the two bands originally intersect near the Fermi energy, strong intrinsic spin-orbit coupling $m_{\bm{k}}$ hybridizes the $d_{yz}$ and $p_y$ orbitals, yielding pronounced orbital mixings near the hybridization gap (Fig. \ref{fig1}(b)). Incorporating this 1T'-monolayer electronic structure with the T$_{\textrm{d}}$-stacking operation gives the effective bilayer Hamiltonian~\cite{Supp}
\begin{align}\label{H-eff-bi}
H\left(\bm{k}\right)=\textrm{diag}\left[H_+\left(\bm{k}\right),H_-\left(\bm{k}\right)\right]
\end{align}
with
\begin{align}
H_\pm\left(\bm{k}\right)=&\begin{pmatrix}
E_{p,\bm{k}}+u & -iv_{\bm{k}}\pm m_{\bm{k}} & t_{pp} & t_{pd} \\
iv_{\bm{k}}\pm m_{\bm{k}} & E_{d,\bm{k}}+u & t_{dp} & t_{dd} \\
t_{pp} & t_{dp} & E_{p,\bm{k}}-u & iv_{\bm{k}}\pm m_{\bm{k}} \\
t_{pd} & t_{dd} & -iv_{\bm{k}}\pm m_{\bm{k}} & E_{d,\bm{k}}-u
\end{pmatrix}
\end{align}
in the basis of $\left[\ket{1,p_y,\pm},\ket{1,d_{yz},\pm}, \ket{2,p_y,\pm},\ket{2,d_{yz},\pm}\right]$. Here, $\ket{\pm}$ denotes the effective spin up and down states, $u$ is the gate-induced electrostatic potential energy, $v_{\bm{k}}$ is the spin-independent coupling between the $p$ and $d$ orbitals, and $t_{ij}$ ($\left\{i,j\right\}=\left\{p,d\right\}$) denotes the interlayer hopping from the $i$ orbital in the top layer to the $j$ orbital in the bottom layer.

Taking the effective Hamiltonian in Eq. \ref{H-eff-bi} to calculate the bare out-of-plane polarization gives~\cite{Supp}
\begin{align}\nonumber\label{P-bare}
P\left(u=0\right)=&-\frac{ed}{2\beta}\sum_{n}\int_{\bm{k}}\textrm{Tr}\left[G\left(i\omega_n,\bm{k}\right)s_0\tau_z\sigma_0\right]\big|_{u=0}\\
\propto&\left(t_{pd}-t_{dp}\right),
\end{align}
with $\int_{\bm{k}}\equiv\frac{1}{\left(2\pi\right)^2}\int d\bm{k}$, $d$ being the distance between the two layers and $G\left(i\omega_n,\bm{k}\right)=1/\left[i\omega_n-H\left(\bm{k}\right)\right]$. Eq. \ref{P-bare} explicitly demonstrates that the stacking-induced asymmetry of the interlayer couplings, $t_{pd}\neq t_{dp}$, drives a spontaneous out-of-plane ferroelectric polarization. Physically, the asymmetric interlayer couplings ($t_{pd} \neq t_{dp}$) first induce an unequal transfer of the $p$- and $d$-orbital electrons between the top and bottom layers, generating a net orbital polarization in the bilayer system (Fig.~\ref{fig1}(c)). Since the atomic $p$ and $d$ orbitals have different on-site energies, the orbital polarization then produces an asymmetric charge distribution across the bilayer~\cite{BTZhou, Supp}, yielding an out-of-plane ferroelectric polarization (Fig.~\ref{fig1}(d)). Consequently, the stacking-induced ferroelectricity is intrinsically tied to the underlying layer polarization of the $p$- and $d$-orbital electrons. Notably, this microscopic mechanism for out-of-plane stacking ferroelectricity shares the same physical origin as the quantum geometric scenario developed in Ref.~\citenum{BTZhou}. Given recent observations that the superconducting state in bilayer T$_{\textrm{d}}$-MoTe$_2$ is suppressed by out-of-plane ferroelectric polarization~\cite{Rhodes, Zizhong}, the intrinsic connection between ferroelectricity and the layer segregation of $p$- and $d$-orbital electrons strongly implies that the superconducting pairing in this material is orbital-dependent.

\emph{Ferroelectric suppression of the intralayer $p$-$d$ orbital pairing}.--- Having established that the stacking-induced ferroelectricity in bilayer T$_{\textrm{d}}$-MoTe$_2$ originates from the asymmetric interlayer couplings $t_{pd} \neq t_{dp}$, we now investigate the specific pairing channels that are vulnerable to suppression by the out-of-plane ferroelectric polarization. As T$_{\textrm{d}}$-MoTe$_2$ is already an intrinsic superconductor in the monolayer limit~\cite{Daniel-Nano}, we focus our analysis on the intralayer pairing channels of the bilayer system and specifically examine pairings in the orbital subspace. Assuming an intralayer short-range electron density-density interaction: $-U\sum_{i=1,2}n_i\left(\bm{r}\right)n_i\left(\bm{r}\right)$, where $n_i\left(\bm{r}\right)$ is the electron density in layer $i$, we calculate the resulting pairing susceptibilities as a function of the interlayer coupling asymmetry $|t_{pd}-t_{dp}|$~\cite{Supp}, as plotted in Fig.~\ref{fig2}(a). Notably, while other types of intralayer pairings exhibit no clear dependence on the interlayer coupling asymmetry, the spin-singlet $p$-$d$ orbital pairing susceptibility $\chi^{\left(3\right)}$ decreases as $|t_{pd}-t_{dp}|$ increases. In good agreement with recent experiments~\cite{Rhodes, Zizhong}, this suppression of pairing susceptibility by ferroelectric polarization firmly establishes the intralayer $p$-$d$ orbital pairing as the most likely pairing symmetry in bilayer T$_{\textrm{d}}$-MoTe$_2$.

To further corroborate the ferroelectric suppression of the intralayer $p$-$d$ orbital pairing, we present the superconducting phase diagram in Fig.~\ref{fig2}(b). The phase diagram is obtained by minimizing the free energy density~\cite{Supp}
\begin{align}\label{SC-ff-01}
f=&\frac{|\Delta|^2}{\bar{U}}+\frac{1}{2}\int_{\bm{k}}\left[\textrm{Tr}H\left(\bm{k}\right)-\frac{1}{\beta}\sum_a\ln\left(1+e^{-\beta E^a_{\bm{k}}}\right)\right]
\end{align}
with respect to the pairing order parameter $\Delta$ at a fixed temperature $T$ and interlayer coupling asymmetry $|t_{pd}-t_{dp}|$. Here, $\bar{U}$ denotes the effective attractive interaction in a unit cell and $E^a_{\bm{k}}$ are the eigen-energies of the corresponding Bogoliubov-de Gennes Hamiltonian~\cite{Supp}. The resulting phase diagram clearly demonstrates the depairing effect of out-of-plane ferroelectric polarization on the intralayer $p$-$d$ orbital pairing, with the critical polarization $P_{\textrm{c}}$ defined by the condition $\Delta\rightarrow0$. This depairing occurs because the out-of-plane polarization spatially segregates the $p$- and $d$-orbital electrons into opposite layers, effectively destroying the intralayer Cooper pairs (Fig.~\ref{fig1}(c)). Crucially, this suppression of intralayer $p$-$d$ orbital pairing by the stacking-induced ferroelectricity naturally accounts for the experimental observation that the superconducting transition temperature decreases sharply from $T_{\textrm{c}}=7$ K in monolayer T$_{\textrm{d}}$-MoTe$_2$ to $T_{\textrm{c}}=2.3$ K in bilayer T$_{\textrm{d}}$-MoTe$_2$~\cite{Rhodes, Zizhong, Daniel-Nano, Wakamura}.

\emph{Landau-Ginzburg theory for the ferroelectric hysteretic switching of superconductivity}.--- While the calculated phase diagram in Fig.~\ref{fig2}(b) shows that the out-of-plane polarization suppresses the intralayer $p$-$d$ orbital pairing, the mechanism underlying the gate-voltage-controlled hysteretic switching of superconductivity remains to be elucidated. To address this, we first note that the bare polarization in Eq. \ref{P-bare} is subject to corrections from the asymmetric charge redistribution across the bilayer~\cite{BTZhou, Jackson}. Because this layer charge imbalance inherently generates an interlayer potential difference, the physical electric polarization must be determined self-consistently via the equation~\cite{Supp}
\begin{align}\label{PP-parallel}
P\left(u,\Delta\right)=2u\epsilon/e,
\end{align}
where $P\left(u,\Delta\right)=-\frac{ed\partial f}{2\partial u}$ can be derived from Eq. \ref{SC-ff-01}, and $\epsilon$ is the electric permittivity. Importantly, Eq. \ref{PP-parallel} establishes an intrinsic connection between $u$ and $t_{pd}-t_{dp}$. Since $u\sim\frac{1}{2}eE_zd$ is directly controlled by an external gate voltage, both $t_{pd}-t_{dp}$ and $P\left(u,\Delta\right)$ become gate-tunable through the self-consistent feedback loop illustrated in Fig.~\ref{fig3}(b) inset. 

After establishing the gate tunable nature of $t_{pd}-t_{dp}$ and $P\left(u,\Delta\right)$, we incorporate both the intralayer $p$-$d$ orbital pairing and the stacking-induced ferroelectricity to construct a phenomenological Landau-Ginzburg free energy density, yielding~\cite{Supp}
\begin{widetext}
\begin{align}\label{Landau-Ginzburg-ff}
f\left(P,\Delta\right)\approx N\left(0\right)\left\langle \frac{T-T_{\textrm{c}0}}{T_{\textrm{c}0}}-\sum_{\nu=\pm,j=1,2}\frac{\tilde{g}_{j,\nu,y}^2\left(\bm{k}\right)+\tilde{g}^2_{j,\nu,z}\left(\bm{k}\right)}{2|\tilde{\bm{g}}_{j,\nu}\left(\bm{k}\right)|^2}D\left(\frac{|\tilde{\bm{g}}_{j,\nu}\left(\bm{k}\right)|}{\pi k_{\textrm{b}}T}\right)\right\rangle_{\bm{k}_{\textrm{F}}}|\Delta|^2+\frac{b_{\textrm{s}}}{2}|\Delta|^4+a_{\textrm{f}}P^2+\frac{b_{\textrm{f}}}{2}P^4-P\cdot E_z.
\end{align}
\end{widetext}
Here, $D\left(\rho\right)=\textrm{Re}\sum_{n=1}^\infty\left(\frac{1}{2n-1+i\rho}-\frac{1}{2n-1}\right)$, $N\left(0\right)$ denotes the electronic density of states at the Fermi level, $T_{\textrm{c}0}$ is the superconducting transition temperature of decoupled layers, and $E_z$ denotes the out-of-plane electric field induced by an external gate voltage. \textcolor{black}{The first two terms, obtained from the Taylor expansion of Eq. \ref{SC-ff-01}, describe the superconducting state, while the last three terms govern the ferroelectric phase transition. The parameters $b_{\textrm{s}}$, $a_{\textrm{f}}$ and $b_{\textrm{f}}$ satisfy $b_{\textrm{s}}>0$, $a_{\textrm{f}}<0$ and $b_{\textrm{f}}>0$, where $a_{\textrm{f}}<0$ guarantees a spontaneous ferroelectric polarization.} Crucially, the coupling between $\Delta$ and $P$ is implicitly embedded in the function $\tilde{\bm{g}}_{j,\nu}\left(\bm{k}\right)$. Since $\tilde{\bm{g}}_{j,\nu}\left(\bm{k}\right)$ depends on the microscopic variables $t_{pd}$, $t_{dp}$ and $u$ that dictate the physical polarization $P$~\cite{Supp}, the gate-tunable $P$ can directly modulate the coefficient of $|\Delta|^2$, thereby tuning the superconducting transition temperature. At sufficiently large gate voltages, the induced $P$ with $t_{pd}-t_{dp}$ exceeds the critical pair-breaking threshold (Fig.~\ref{fig3}(b)) and destroys the superconductivity.

\color{black}

With the established Landau-Ginzburg formalism in Eq. \ref{Landau-Ginzburg-ff}, we are ready to elucidate the mechanism behind the hysteretic switching of superconductivity. At zero external field ($E_z=0$), combining Eq. \ref{PP-parallel} and \ref{Landau-Ginzburg-ff} yields the density plot of $f\left(P,\Delta\right)$ shown in Fig.~\ref{fig3}(a), where the global minimum at finite $\Delta$ and $P$ manifests the coexistence of intralayer $p$-$d$ orbital pairing and ferroelectric polarization. This coexistence requires $P_{\textrm{c}}>P_{\textrm{r}}$. Under a cyclic gate sweep, the ferroelectric order exhibits a hysteresis schematically shown in Fig.~\ref{fig3}(b). Along the hysteresis path, the ferroelectric polarization reverses direction at the coercive field $E_{\textrm{c}}$ and converges to the saturated polarization $P_{\textrm{s}}$. Crucially, transport experiments in bilayer T$_{\textrm{d}}$-MoTe$_2$ show that the switching of superconductivity and ferroelectric reversal occur concurrently at $E_{\textrm{c}}$~\cite{Rhodes, Zizhong}, establishing that the critical polarization satisfies $P_{\textrm{c}}<P_{\textrm{s}}$. Therefore, under the condition $P_{\textrm{r}}<P_{\textrm{c}}<P_{\textrm{s}}$, the hysteretic switching of superconductivity naturally arises as the polarization hysteretically exceeds $P_{\textrm{c}}$ in the hysteresis loop (Fig.~\ref{fig3}(b) and (c)).

Based on this mechanism, we simulate the ferroelectric hysteresis of the intralayer, $p$-$d$ orbital pairing via minimizing $f\left(P,\Delta\right)$ with respect to $P$ and $\Delta$ across a range of applied electric fields $E_z$. Combining Eq. \ref{PP-parallel} and \ref{Landau-Ginzburg-ff}, we obtain both the hysteresis loop of pairing order parameter and the corresponding temperature evolution, shown in Fig.~\ref{fig3}(c) and \ref{fig3}(d), respectively. Crucially, the pairing hysteresis between $\Delta=0$ and finite $\Delta$ in Fig.~\ref{fig3}(c) maps directly onto the experimentally observed transport hysteresis between finite resistance and superconducting states~\cite{Rhodes, Zizhong}. Notably, Fig.~\ref{fig3}(d) captures the dramatic enhancement of $T_{\textrm{c}}$ as ferroelectric polarization approaches reversal, consistent with recent experiments on bilayer T$_{\textrm{d}}$-MoTe$_2$~\cite{Rhodes}. In realistic devices, substrate and contact asymmetries~\cite{Zaiyao, Hunt, Pablo2, Rhodes, Zizhong} usually introduce an internal field offset $\delta E$, which shifts $E_z$ to $E_z+\delta E$ and gives rise to an asymmetric hysteresis loop (see Fig.~S4 in the Supplemental Material~\cite{Supp}).

\emph{Implications of intralayer $p$-$d$ orbital pairing}.--- For the superconducting bilayer T$_{\textrm{d}}$-MoTe$_2$, \textcolor{black}{we have established the intralayer, $p$-$d$ orbital pairing as the physical origin of ferroelectric hysteretic switching of superconductivity. This naturally raises the question of how to experimentally verify} our scenario. A key consequence of this intralayer, $p$-$d$ orbital pairing is that the pairing gap on the Fermi surface is highly non-uniform. As the $p$-$d$ orbital pairing relies on the inter-orbital hybridization, the pairing gap is maximized at regions of strong inter-orbital hybridization but gets significantly suppressed elsewhere (Fig. \ref{fig4}(a)). Therefore, measuring the superconducting gap anisotropy via high-resolution angle-resolved photoemission spectroscopy~\cite{ZXShen01,ZXShen02} serves as a direct test of the $p$-$d$ orbital pairing we proposed for bilayer T$_{\textrm{d}}$-MoTe$_2$.

Another physical implication is that the system can stabilize a PDW as the Fermi level is gate-tuned away from the strong orbital hybridization regime. In bilayer T$_{\textrm{d}}$-MoTe$_2$, the superconducting transition temperature is observed to reach the maximum when electron and hole carrier densities are close to compensation~\cite{Rhodes, Zizhong, Daniel-Nano}. This aligns well with our $p$-$d$ orbital pairing scenario, as the optimal hybridization between the $p$ and $d$ orbitals occurs within the compensated regime~\cite{Supp}. When the Fermi level shifts away from the strong hybridization regime, the Fermi pockets from the $p$ and $d$ orbitals gradually separate in $\bm{k}$ space, leading to a decrease in the superconducting $T_{\textrm{c}}$ (Fig. \ref{fig4}(b)). Remarkably, the intralayer $p$-$d$ orbital pairing interaction can even enforce a finite-momentum modulation of the $p$-$d$ orbital pairing and stabilize a PDW state in the $\mu$-$T$ phase diagram shown in Fig. \ref{fig4}(b)~\cite{Supp}. Consequently, identifying a gate-induced PDW in bilayer (or monolayer) T$_{\textrm{d}}$-MoTe$_2$ would provide a definitive validation of our $p$-$d$ orbital pairing scenario.

\emph{Discussions}.--- In this work, we have developed an intralayer, $p$-$d$ orbital pairing scenario to explain the ferroelectric hysteretic switching of superconductivity in bilayer T$_{\textrm{d}}$-MoTe$_2$. Crucially, we show that this ferroelectric hysteretic behavior originates from three coupled factors: (i) the field-induced hysteresis of electric polarization; (ii) the polarization-induced modulation of superconducting transition temperature; (iii) the hysteretic switching criterion $P_{\textrm{r}}<P_{\textrm{c}}<P_{\textrm{s}}$. In realistic 2D devices, structural fluctuations, local strain and disorder introduce spatial modulations to the ferroelectric polarization, thereby transforming the ideal hysteresis loop (Fig.~\ref{fig3}(b)) into the highly asymmetric profile observed experimentally~\cite{Rhodes, Zizhong}. In the atomically thin 2D material, long-range superconducting order is susceptible to thermal phase fluctuations~\cite{Mermin}, but quasi-long-range phase coherence is retained below the Berezinskii-Kosterlitz-Thouless (BKT) transition temperature~\cite{Berezinskii, Kosterlitz}. Since the superfluid stiffness scales with $\Delta^2$~\cite{Tinkham}, the polarization-induced modulation of $\Delta$ maps directly onto a modulation of phase rigidity. Consequently, full consideration of the BKT physics simply remaps the mean-field pair-breaking boundary onto the corresponding BKT transition. As long as the critical polarization that induces the BKT transition lies within the polarization hysteresis loop, the hysteretic switching of superconductivity is guaranteed to occur (see Fig. S5 in the Supplemental Material). 

\color{black}

Our theory establishes a paradigm for the ferroelectric hysteretic switching of superconductivity, rooted in the interplay between intralayer, inter-orbital pairing and a layer-polarized charge distribution. This framework is generalizable to other ferroelectric superconductors with more complex spatial structures~\cite{FangHong, J-Wu, Yanwu, Zhenyu02, Zhenyu03, Rowley} by considering the coupling of inter-orbital pairing to their specific charge polarization profiles. Intriguingly, the recent observation of orbital magnetic hysteresis persisting into the superconducting states in rhombohedral graphene~\cite{Tonghang} and twisted bilayer MoTe$_2$~\cite{Tingxin01} exhibits a formal similarity to the ferroelectric hysteresis studied here. Consequently, our phenomenological theory of ferroelectric-superconducting coupling may provide a pathway toward understanding the coupling between orbital magnetism and superconductivity in these systems, which are promising candidates for hosting chiral superconductivity.

\emph{Acknowledgements}.--- W.-Y. H. thanks Patrick A. Lee, Tao Xiang and Yoichi Yanase for helpful discussions. W.-Y.H. acknowledges the support from the National Natural Science Foundation of China (No. 12304200, No. 12674202), the BHYJRC Program from the Ministry of Education of China (No. SPST-RC-10), the Shanghai Rising-Star Program (24QA2705400), the Science and Technology Commission of Shanghai Municipality (No. 24LZ1401000) and the start-up funding from ShanghaiTech University.

\end{document}